\documentclass[prl,aps,amsmath,amssym,amsfonts,floats,floatfix,twocolumn]{revtex4-2}

\usepackage{epsfig} 
\usepackage{psfrag}
\usepackage{braket}
\usepackage{textcomp}
\usepackage{graphicx}
\usepackage{subfigure}
\begin{document}

%\title{Gauging frustration in bundles of twisted filaments: from symmetry to elasticity}

\title{Mechanics of metric frustration in contorted filament bundles:  \\ From local symmetry to columnar elasticity}

\author{Daria W. Atkinson}
\affiliation{Department of Physics and Astronomy, University of Pennsylvania}
\email{atkinsod@sas.upenn.edu, she/her or they/them}
\author{Christian D. Santangelo}
\affiliation{Department of Physics, Syracuse University}
\author{Gregory M. Grason}
\affiliation{Department of Polymer Science and Engineering, University of Massachusetts Amherst}
\begin{abstract}
Bundles of filaments are subject to geometric frustration:  certain deformations (e.g. bending while twisted) require longitudinal variations in spacing between filaments. While bundles are common---from protein fibers to yarns---the mechanical consequences of longitudinal frustration are unknown. We derive a geometrically-nonlinear formalism for bundle mechanics, using a gauge-like symmetry under {\it reptations} along filament backbones. We relate force balance to orientational geometry and assess the elastic cost of frustration in twisted toroidal bundles.
\end{abstract}
\maketitle
%We still need citations here!
Elastic zero modes are a ubiquitous feature of soft materials, from mechanical metamaterials~\cite{mao_soft_2010,sun_surface_2012} to liquid crystal elastomers \cite{warner_soft_1994}.  Such systems can undergo large deformations with minimal strain, as geometrically coupled rotations and translations preserve local spacing between microscopic constituents.  The smectic and columnar liquid crystalline phases provide paradigmatic examples of zero modes in soft elastic systems, permitting relative ``sliding'' of 2D layers and 1D columns, respectively.  The zero-cost sliding displacements of smectic and columnar phases are characteristic of a much broader class of laminated and filamentous structures, ranging from multi-layer graphene materials \cite{xu_ultrastrong_2013} and stacked paper \cite{poincloux_what_2020} to biopolymer bundles \cite{hud_cryoelectron_2001,leforestier_structure_2009}, nanotube yarns \cite{zhang_multifunctional_2004}, wire ropes \cite{costello_theory_1990}.

\begin{figure}
\centering
\includegraphics[width=0.95\columnwidth]{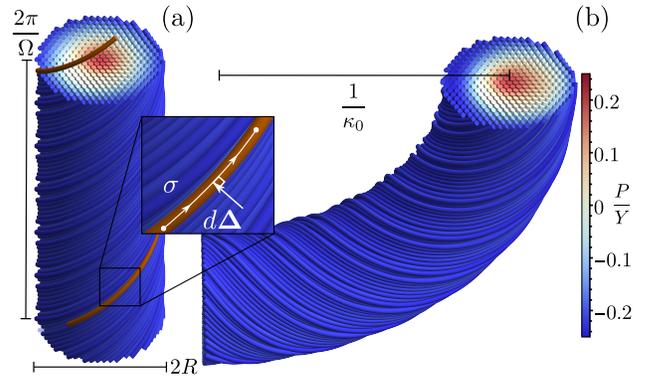}
\caption{In~(a), an equilibrium twisted bundle, with $\Omega R = 1$ and 2D Poisson ratio $\nu = 0.8$, colored by the local pressure.   {\it Reptation} of the orange filament by $\sigma$ along its contour leaves local separation $d\boldsymbol{\Delta}$ unchanged. In~(b), a twisted-toroidal bundle found by bending the same bundle to $\kappa_0  = 0.2/R$ and optimizing inter-filament elastic cost.  }\label{fig:fig1}
\end{figure}

While there are well established frameworks which capture the geometric nonlinearities of smectic liquid crystals (i.e., the strain tensor accurately describes arbitrarily large and complex deformations) \cite{grinstein_nonlinear_1982}, no such framework exists for columnar and filamentous materials. The orientations of column backbones impose constraints on inter-filament spacing, generating rich modes of geometric frustration without counterpart in smectic liquid crystals.
In the simplest non-trivial case of helical bundles, predictions from a minimally non-linear approximation of columnar elasticity~\cite{grason_defects_2012} and tomographic analysis of elastic filament bundles~\cite{panaitescu_measuring_2017} show that twist in straight bundles gives rise to non-uniform inter-filament stress and spacing in transverse sections (see Fig~\ref{fig:fig1}a).
Except for the restrictive classes of straight, twisted bundles~\cite{grason_colloquium_2015} and twist-free developable domains~\cite{bouligand_geometry_1980, kleman_developable_1980}, bundle textures also generate {\it longitudinal} frustration, requiring local spacings to vary {\it along} a bundle \cite{atkinson_constant_2019}.  Although deformations that introduce longitudinal frustration are the rule rather than the exception---for example, wire ropes or toroidal biopolymer condensates are both {\it twisted} and {\it bent} (e.g. Fig.~\ref{fig:fig1})---existing frameworks of columnar elasticity fail to capture this effect.

In this Letter, we develop a fully geometrically non-linear Lagrangian elasticity theory of columnar materials, which completely captures the interplay between orientation, and both lateral and longitudinal frustration of inter-filament spacing.  We construct this theory by imposing a gauge-like local symmetry under {\it reptations}, deformations that slide filaments along their contours without changing the inter-filament spacing.  The resultant equilibrium equations point to the role geometrical measures of non-equidistance play in bundles' mechanics.  Within this framework, we compute the energetic costs of longitudinal frustration in twisted, toroidal bundles, and give evidence that 1)  optimal configurations generically incorporate splay and 2)  the bending cost that derives from non-uniform compression depends non-monotonically on pretwist.

To construct the elastic theory, we divide space into points on curves (i.e. filament backbones) labeled by two coordinates: $\mathbf{v}$, a 2D label of filaments; and $s$, a length coordinate along filaments. Hence, the location of each point in the bundle is described by a function $\mathbf{r}(s,\mathbf{v})$, with $\partial_s \mathbf{r}(s,\mathbf{v})$  parallel to the tangent vector $\mathbf{t} = {\partial_s \mathbf{r}}/{|\partial_s \mathbf{r}|}$. Because they lack positional order along their backbone curves, filament bundles and columnar liquid crystals have a family of continuous zero modes, corresponding to {\it reptations} (Fig. \ref{fig:fig1}),
\begin{equation}\label{eq:symm}
\mathbf{r}'(s,\mathbf{v}) = \mathbf{r}(s+\sigma(s,\mathbf{v}),\mathbf{v}).
\end{equation}
We assume that changes in local spacing can be described by a hyper-elastic energy density function ${\cal W}$, that depends only on the deformation gradient~\cite{ciarlet_introduction_2005}.

To account for reptation symmetry, we demand that ${\cal W}$ depend on a modified deformation gradient, which transforms as a scalar under $\sigma(s,\mathbf{v})$, depends solely on the deformation, $\mathbf{r}$, of the material itself, and recovers the well-established 2D elasticity of developable~\cite{bouligand_geometry_1980,kleman_developable_1980} bundles.
Specifically, we construct a covariant derivative $D_I \mathbf{r} = \nabla_I \mathbf{r} - \mathbf{A}_I$, where $\nabla_I$ is the usual covariant derivative on tensors in the material space and $\mathbf{r}$ determines $\mathbf{A}_I$, such that if two configurations, $\mathbf{r}$ and $\mathbf{r}'$, are related by Eq. (\ref{eq:symm}), then $D_I \mathbf{r} = D'_I \mathbf{r}' $.  In order for $D_I\mathbf{r}$ to be reptation invariant, we must have that $- \mathbf{A}'_I + \mathbf{A}_I = \nabla_I \sigma \partial_s \mathbf{r}$. Therefore, $- \mathbf{A}'_I + \mathbf{A}_I = - (\mathbf{t} \cdot \nabla_I \mathbf{r}') \mathbf{t} + (\mathbf{t} \cdot \nabla_I \mathbf{r}) \mathbf{t}$.  To construct an elastic theory of columnar materials, we set $ \mathbf{A}_I=  (\mathbf{t} \cdot \nabla_I \mathbf{r}) \mathbf{t}$, which is manifestly reptation invariant but also leads to a deformation gradient that only measures deformations {\it transverse} to the local backbones,
\begin{equation}\label{eq:covar}
D_I \mathbf{r} \equiv \nabla_I \mathbf{r} - (\mathbf{t} \cdot \nabla_I \mathbf{r}) \mathbf{t}.
\end{equation}
Notably, for two nearby filaments at ${\bf r}({\bf v})$ and ${\bf r}({\bf v}+d{\bf v})$ in material coordinates, it is straightforward to show that the covariant derivative gives the local {\it distance of closest approach} $d {\bf \Delta} = d v^I ~D_I \mathbf{r}$, for which $d {\bf \Delta} \cdot {\bf t} =0$ (see again inset of Fig. \ref{fig:fig1}a). As shown explicitly in the Appendix, this covariant deformation gradient captures the standard 2D deformation gradients of developable domains (i.e. parallel arrays).

From this deformation gradient, we construct an effective metric $g^{\text{eff}}_{IJ} = D_I \mathbf{r} \cdot D_J \mathbf{r}$, which is naturally invariant under rotations of $\mathbf{r}$, and which encodes the metric inherited by the local 2D section transverse to the filaments in the bundle~\footnote{Equidistant bundles where components of $g^{\text{eff}}_{IJ}$ are uniform along $s$ are known as Riemannian fibrations, see e.g. ~\cite{oneill_fundamental_1966,gromoll_metric_2009}, and the transverse structure can be mapped on to single 2D base space.  Non-equidistant bundles correspond to Sub-Riemannian generalizations~\cite{agrachev_sub-riemannian_2019} where the inherited metric measures distances in the 2-planes perpendicular to ${\bf t}$ at each point.}.  Because $D_s \mathbf{r} =0$, the effective metric only has components for $2 \times 2$ block $I, J \neq s$, which we denote using index notation $\alpha$, $\beta \in \{1,2 \}$.  With these definitions, we construct the Green-Saint-Venant strain tensor, \cite{ciarlet_introduction_2005,efrati_elastic_2009,dias_programmed_2011}
\begin{equation}
\label{eq: epsilon}
\epsilon_{\alpha \beta} = \tfrac{1}{2} \big [D_\alpha \mathbf{r} \cdot D_\beta \mathbf{r} - g_{\alpha \beta}^{\text{tar}} \big ],
\end{equation}
where $g_{\alpha \beta}^{\text{tar}}$ is the {\it target metric} corresponding to strain-free state, which for this Letter, we take to be developable with uniform spacing, so $g_{\alpha \beta}^{\text{tar}} = \delta_{\alpha \beta}$.   For weak deflections from the uniform parallel state, Eq. (\ref{eq: epsilon}) reduces to the small-tilt approximation to the non-linear columnar strain~\cite{selinger_hexagonal_1991,grason_chirality_2007,grason_defects_2012}, which captures the lowest-order dependence of spacing on orientation (see Appendix).

Assuming that strains are small though deformations may be large, the Hookean elastic energy takes the usual form,
\begin{equation}
{\cal E}_s =\int dV ~ {\cal W} (\epsilon)   = \frac{1}{2} \int dV S^{\alpha \beta} \epsilon_{\alpha \beta},
\label{eqn:straineng}
\end{equation}
where $S^{\alpha \beta} =  \tfrac{\partial {\cal W}}{\partial \epsilon_{\alpha \beta} } =C^{\alpha \beta \gamma \delta} \epsilon_{ \gamma \delta}$ is the nominal stress tensor, and $C^{\alpha \beta \gamma \delta}$ is a tensor of elastic constants which depends on both the crystalline symmetries of the underlying columnar order and the target metric, $g_{\alpha \beta }^{\text{tar}}$~\footnote{We raise and lower indices with $g^{\alpha \beta \text{tar}}$, the inverse of $g_{\alpha \beta}^{\text{tar}}$, rather than $g^{\alpha \beta \text{eff}}$, as a matter of convenience, and note that the difference in the elastic energy is higher order in the strain tensor~\cite{dias_programmed_2011}.}.   Energetics of columnar materials also include other gauge-invariant costs, including the Frank-Oseen orientational free energy and the cost of local density changes {\it along} columns~\footnote{The symmetry under local reptations in this theory is unlike a related treatment of both the nematic to smectic-A transition \cite{de_gennes_analogy_1972,lubensky_gauge_1982} and its columnar analog \cite{giannessi_frank_1986,kamien_iterated_1995,kamien_liquids_1996} in which the gauge symmetry of a density-wave is explicitly broken by, e.g., the splay elasticity of the underlying nematic order.}.  Here, for clarity, we focus only on the energetics of columnar strain and detail the combined effects of other contributions elsewhere \cite{atkinson_fibers_2020}. Given this gauge-invariant formulation of the columnar strain energy, we first  
illustrate the mechanical effects of orientational geometry on local forces.   This follows from the bulk Euler-Lagrange equations of Eq.~\eqref{eqn:straineng} (see Appendix for a complete derivation),
\begin{equation}
\label{eqn:el}
\frac{\delta \mathcal{E}_s}{\delta {\bf r} } =  -  \nabla_\alpha \big( S^{\alpha \beta} D_\beta \mathbf{r} \big )+  \nabla_s \Big( S^{\alpha \beta} D_\beta \mathbf{r} \frac{\mathbf{t} \cdot \nabla_\alpha \mathbf{r}}{|\nabla_s \mathbf{r}|} \Big ). 
\end{equation}
The bulk terms represent body forces generated by the columnar strains, which must be balanced by other internal stresses or external forces.  To cast them in a more geometrical light, we consider separately the components tangential and perpendicular to ${\bf t}$, $F_\parallel$ and ${\bf F}_\perp$, respectively.  

Making use of the identity, ${\bf t} \cdot \nabla_\alpha D_\beta {\bf r} = - \partial_\alpha {\bf t} \cdot \partial_\beta {\bf r}$, the tangential forces can be recast simply as
\begin{equation}
F_\parallel = S^{\alpha \beta} h_{\alpha \beta} ,
\label{eqn:betteryetelbulkpar}
\end{equation}
where 
\begin{align}
h_{\alpha \beta} &= \tfrac{1}{2} \big [ \partial_\alpha \mathbf{t} \cdot \partial_\beta \mathbf{r} + \partial_\beta \mathbf{t} \cdot \partial_\alpha \mathbf{r}  \\
&-  \frac{\mathbf{t} \cdot \partial_\alpha \mathbf{r}}{|\partial_s \mathbf{r}|}\partial_s \mathbf{t} \cdot \partial_\beta \mathbf{r} - \frac{\mathbf{t} \cdot \partial_\beta \mathbf{r}}{|\partial_s \mathbf{r}|}\partial_s \mathbf{t} \cdot \partial_\alpha \mathbf{r} \big], \nonumber
\end{align}
is the {\it convective flow tensor} that measures longitudinal variations in inter-filament spacing~\cite{atkinson_constant_2019}.  Just as the second fundamental form measures gradients of a surface's normal vector \cite{millman_elements_1977}, $h_{\alpha \beta}$ measure the symmetric gradients of $\mathbf{t}$ in its normal plane (i.e. its trace is the splay of filament tangents).

\begin{figure}
\includegraphics[width=0.85\columnwidth]{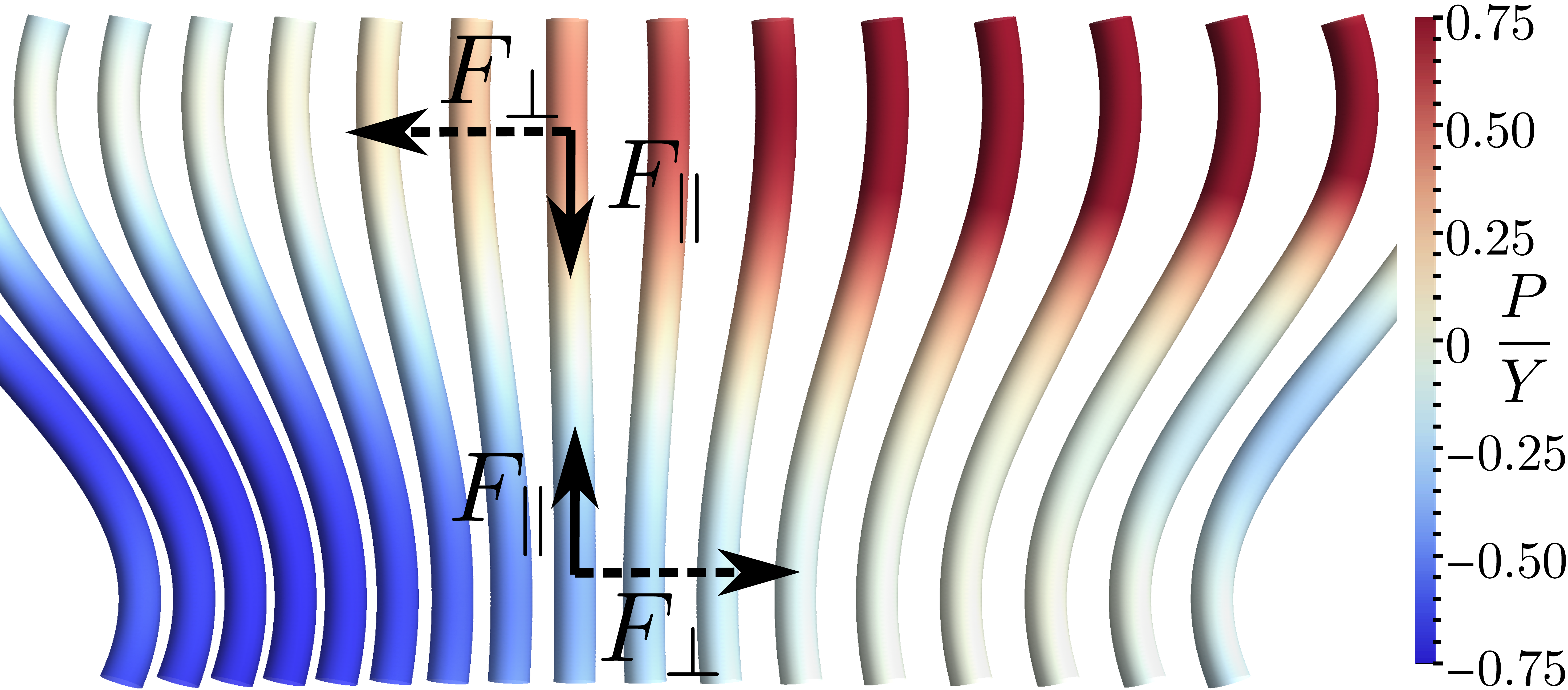}
\caption{A 2D section of filament bundle colored by the local pressure, with regions under compression in blue and those under extension in red.  The forces parallel and tangent to $\mathbf{t}$ are shown at two points, and push material points to regions of vanishing stress.}
\label{fig:tangentforcebalance}
\end{figure}

Here, we see that tangent forces couple non-equidistance to the stress tensor much like the Young-Laplace law couples in-plane stresses to normal forces in curved membranes \cite{efrati_elastic_2009}. This analogy becomes exact for zero twist textures: when $\mathbf{t} \cdot (\nabla \times \mathbf{t}) = 0$, filaments can be described by a set of surfaces normal to $\mathbf{t}$.  In this case, it is possible to choose coordinates so that $\partial_\alpha {\bf r} \cdot {\bf t} =0$, and tangential forces give the Young-Laplace force normal to each surface, with $h_{\alpha \beta}$ reducing to their second fundamental form.  This illustrates the intuitive notion, shown schematically in Fig.~\ref{fig:tangentforcebalance},  that columnar strain generates tangential body forces that push material points {\it towards} lower-stress locations in the array.

The bulk components of Eq.~\eqref{eqn:el} perpendicular to $\mathbf{t}$ give the transverse force
\begin{align}
\mathbf{F}_\perp &= - D_\alpha \big [ S^{\alpha \beta} D_\beta \mathbf{r}\big ]  +  D_{s} \big [ S^{\alpha \beta} \frac{\mathbf{t} \cdot \partial_\alpha \mathbf{r}}{|\partial_s \mathbf{r}|} D_\beta\mathbf{r}\big ]
\label{eqn:betteryetelbulkperp}.
\end{align}
This form captures the divergence of stress in the planes perpendicular to backbones.  The second term accounts for the corrections arising from material derivatives that lie along the backbone, such as twisted textures, when $\partial_\alpha {\bf r} \cdot {\bf t} \neq 0$.  Thus, the longitudinal derivatives in $\mathbf{F}_\perp$ are needed to capture transverse mechanics of even equidistant twisted bundles beyond the lowest order geometric non-linearity~\cite{grason_defects_2012}.

We now illustrate the energetics of longitudinal frustration by considering a prototypical non-equidistant geometry: twisted toroidal bundles.   Motivated in large part  by the morphologies of condensed biopolymers~\cite{hud_cryoelectron_2001,leforestier_structure_2009}, theoretical models of twisted toroids have focused on their orientational elasticity costs~\cite{kulic_twist-bend_2004, charvolin_geometrical_2008, koning_saddle-splay_2014}, ignoring the unavoidable frustration of inter-filament spacing in this geometry.  While satisfying force balance in non-equidistant bundles requires physical ingredients beyond the columnar strain energy, which we consider elsewhere \cite{atkinson_fibers_2020}, for the purposes of this Letter we take advantage of the full geometric-nonlinearity of Eq.~\eqref{eqn:el} to explore the specific costs of longitudinal gradients in spacing required by simultaneous twist and bend.  

We construct twisted toroids from equilibrium twisted helical bundles of radius $R$ and constant pitch, ${2 \pi}{/\Omega}$ by bending them such that their central curve ${\bf r}_0$ is deformed from a straight line into a circle of radius $\kappa_0^{-1}$ (see Fig.~\ref{fig:fig1}b).    We then define perturbative displacements $\mathbf{r}(s+ \delta s, \rho + \delta \rho, \phi + \delta \phi)  \simeq \mathbf{r}_0 + \rho \hat{\rho} + \partial_s\mathbf{r}^{(0)} \delta s + \delta \rho \hat{\rho} + \delta \phi \hat{\phi}$ relative to the bent, pre-twisted bundles, where $\rho$ and $\phi$ describe the (Eulerian) distance from the central curve and the angular position relative to its normal in the plane perpendicular to its tangent ${\bf t}_0$, and where $\partial_s\mathbf{r}^{(0)} = {\bf t}_0 + \Omega \rho \hat{\phi}$.  The small-$\rho$ limit of the force balance equations for the strain energy motivates the following displacements
\begin{equation}
\begin{pmatrix}
\delta s \\
\delta \rho \\
\delta \phi
\end{pmatrix}
=
\begin{pmatrix}
a_s \Omega \kappa_0 \rho^3 \sin{\phi} \\
a_\rho \Omega^2 \kappa_0 \rho^4 \cos{\phi} \\
a_\phi \Omega^2 \kappa_0 \rho^3 \sin{\phi}
\end{pmatrix}
\label{eqn:ansatzdef}
\end{equation}
where $a_s$, $a_\rho$, and $a_\phi$ are variational parameters.  Notably, to linear order in curvature, these parameterize the``almost equidistant'' ansatzes  considered previously,  including splay-free ($\text{tr}(h)=0$)~\cite{kulic_twist-bend_2004} configurations and $\text{det}(h)=0$ \cite{atkinson_constant_2019} ansatzes.  

We expand the energy to quadratic order in $\kappa_0$, holding the center of area at $\mathbf{r}_0$, which constrains $a_s = a_\rho - a_\phi$, then minimize Eq.~\eqref{eqn:straineng} with respect to the displacements for a given $\Omega R$.  Examples of the distribution of pressure $P=S^{\alpha}_{ \alpha}/2$ are shown in Fig.~\ref{fig:fig1}b.  Relative to the axisymmetric pressure induced by helical twist in the straight bundle, bending into a twisted toroid requires bunching (spreading) of the filaments at the inner (outer) positions in the toroid, leading to a polarization of the pressure towards the normal.

\begin{figure}
\centering
\includegraphics[width=0.95\columnwidth]{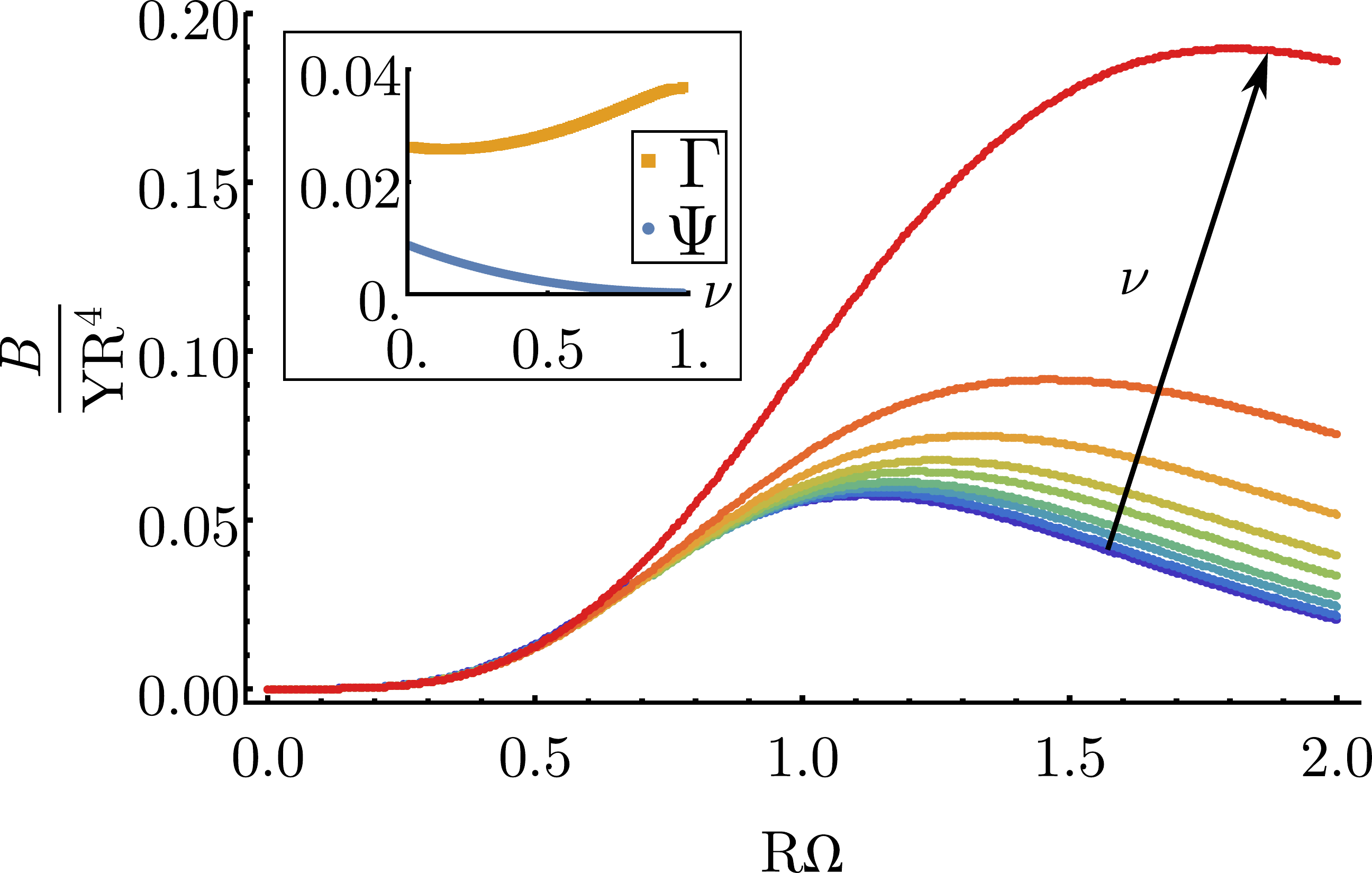}
\caption{The effective bending modulus $B$ which results from the frustration of constant spacing in twisted-toroidal bundles.  The inset shows dimensionless measures of mean splay ($\Psi$) and biaxial splay ($\Gamma$) defined in the text, for fixed twist $\Omega R = 0.8$ for a range of 2D Poisson ratios, from $\nu=0.2$ to 1.0.  Approaching incompressiblity ($\nu \to 1$), $\Psi$ goes to zero, maintaining uniform area-per-filament at the expense of expense of markedly increased strain energy.}
\label{fig:bundleenergetics}
\end{figure}

Because bending and twisting of bundles introduces longitudinal strain variation, bending pre-twisted bundle introduces additional stresses whose elastic cost can be characterized by an effective bending stiffness $B$, defined by $E_\text{bend} = \tfrac{B}{2} \int ds \kappa_0^2$, which derives purely from columnar strain, rather than intra-filament deformations.  Fig.~\ref{fig:bundleenergetics} shows that longitudinal frustration leads to a bending cost that increases with small twist as $B \sim  (\Omega R)^4$, but eventually gives way to remarkable non-monotonic behavior at large pre-twist.  We note further that the bending cost grows with the 2D Poisson ratio $\nu$ of the columnar array, highlighting the importance of local compressional deformations in optimal twisted toroids.

We analyze the optimal modes of deformation via the convective flow tensor, in particular, the~trace ${\rm tr} (h)$ (splay) and deviatoric components $h^{\text{dev}} = h_{\alpha \beta} - {\rm tr} (h) \delta_{\alpha \beta}/2$ (biaxial splay) \cite{selinger_interpretation_2018}, which characterize longitudinal gradients of dilatory and shear stress in the columnar array.  In contrast to a heuristic view that optimal packings should favor the uniform area per filament of splay-free textures, the inset of Fig.~\ref{fig:bundleenergetics} instead shows that optimal toroids incorporate a mixture of both splay and biaxial splay where we define the respective measures of average splay and biaxial splay,  $\Psi \equiv \tfrac{1}{\kappa_0^2 V} \int dV \text{tr}(h)^2$ and $\Gamma \equiv \tfrac{1}{\kappa_0^2 V} \int dV \text{tr}(h_{\text{dev}}^2)$.  Only in the incompressible limit, as $\nu \to 1$, does the splay vanish, and only at the expense of additional biaxial splay and energetic cost, implying counterintuitively that splayed textures are in fact energetically favorable in longitudinally frustrated twisted toroids.  Indeed, the energetic preference for splay in non-equidistant bundles, can be traced to force balance conditions in this geometry~\cite{atkinson_fibers_2020}.

In summary, we have shown that gauge-theoretic principles underlie the geometrically-nonlinear theory of columnar elasticity, providing a means to quantify the cost of longitudinal frustration in the mechanics of bundles.  Unlike phase field models of nonlinear elasticity (such as \cite{chaikin_principles_1995,stenull_phase_2009}), this description depends neither on the presence of a planar reference state, nor presupposes uniform crystalline order, allowing us to both accommodate the effective curvature of bundles of constant pitch helices \cite{bruss_non-euclidean_2012}, and providing a natural generalization to arbitrary target metrics \cite{efrati_elastic_2009}.  Finally, because this approach to elasticity relies only on the existence of local, continuous zero modes, we note that it can be generalized to other liquid crystals, like smectics, and anticipate that it may have applications beyond liquid crystals, including mechanical metamaterials.

\begin{acknowledgments}
The authors gratefully acknowledge useful discussions with R. Kusner, S. Zhou, and B. Davidovitch. 
M. Dimitriyev, D. Hall, and R. Kamien provided feedback on an early draft of this manuscript.  This work was supported by the NSF under grants DMR-1608862 and NSF DMR-1822638;
D.A. received additional funding from the Penn Provost's postdoctoral fellowship and NSF MRSEC DMR-1720530.
\end{acknowledgments}

\bibliography{BundleSources}

\clearpage

\onecolumngrid
\appendix

\section{The form of the deformation gradient}
To construct the covariant derivative in Eq.~(2), we demand, in addition to reptation symmetry and dependence only on gradients of the deformation, that $D_I \mathbf{r}= \nabla \mathbf{r} - \mathbf{A}_I $ reduce to the usual 2D deformation gradient for the well studied {\it developable} bundles \cite{kleman_columnar_1982,grason_frustration_2013}, where the cross-sectional geometry is Euclidean \cite{bouligand_geometry_1980,kleman_developable_1980}. This constrains the value of $\mathbf{A}_I =(\mathbf{t} \cdot \partial_I \mathbf{r}) \mathbf{t}$ in Eq.~(2). Here, we illustrate that this reduces to the expected 2D planar elasticity of developable bundles.
%It might be nice to have a not-group reference that explicitly treats the 2d elasticity of not straight developable domains, but I'm, having trouble finding one.

In a developable bundle, all curves are everywhere to a common set of planes and therefore, share a common tangent in those planes.  Choosing a curve in the bundle $\mathbf{r}_0(s)$ with tangent vector $\hat{T}_0$, this condition requires that ${\bf t}(s, \vec{v}) = \hat{T}_0 (s)$.  Introducing a ``twist-free'', right-handed, orthonormal frame $\{\hat{e}_1(s), \hat{e}_2(s), \hat{T}_0(s) \}$, 
\begin{align}
\partial_s \hat{T}_0 &= \kappa_1 \hat{e}_1 + \kappa_2 \hat{e}_2 \nonumber \\
\partial_s \hat{e}_1 &= - \kappa_1 \hat{T}_0  \\
\partial_s \hat{e}_2 &= - \kappa_2 \hat{T}_0. \nonumber
\end{align}
We can now see that a deformation $\mathbf{r}$ yields a developable bundle (up to reptations) when it can be written as
\begin{equation}
\label{eq: develop}
\mathbf{r}(s, \vec{v}) = \mathbf{r}_0(s) + w_1(\vec{v}) \hat{e}_1(s) + w_2(\vec{v}) \hat{e}_2(s),
\end{equation}
where $w_1$ and $w_2$ are deformation fields depending only on the orthogonal ($\vec{v}$) position, as the tangent field $\mathbf{t} = \partial_s \mathbf{r}/|\partial_s \mathbf{r}|$ is independent of $v_1$ and $v_2$.

The most generic form of the covariant deformation gradient invariant under reptations is
\begin{equation}
\tilde{D}_I \mathbf{r} = \partial_I \mathbf{r} - \tilde{\mathbf{A}}_I \big[\nabla \mathbf{r}\big] - (\mathbf{t} \cdot \partial_I \mathbf{r}) \mathbf{t},
\end{equation}
where $\tilde{\mathbf{A}}_I \big[\nabla \mathbf{r}\big]$ is an unknown function of $\nabla \mathbf{r}$ (i.e. a potential non-zero  value of $\mathbf{A}_I -(\mathbf{t} \cdot \partial_I \mathbf{r}) \mathbf{t}$).  From eq. (\ref{eq: develop}) we find that
\begin{align}
\tilde{D}_s \mathbf{r} &= - \tilde{\mathbf{A}}_s \\
\tilde{D}_{v_1} \mathbf{r} &= \partial_{v_1} w_1~\hat{e}_1 + \partial_{v_1} w_2~\hat{e}_2 - \tilde{\mathbf{A}}_{v_1} \\
\tilde{D}_{v_2} \mathbf{r} &= \partial_{v_2} w_1~\hat{e}_1 + \partial_{v_2} w_2~\hat{e}_2  - \tilde{\mathbf{A}}_{v_2}.
\end{align}
Namely, $-\tilde{\mathbf{A}}_s$ is the only term remaining in the $s$ component, while in the $v_1$ and $v_2$ components, $\tilde{\mathbf{A}}_{v_\alpha}$ is subtracted from the standard 2D deformation gradient in the planes normal to $\hat{T}_0$.  Hence, in order that strains recover the elastic distortions of transverse distances in the columnar structure, for developable structure we must have $\tilde{\mathbf{A}}_{v_1} = \tilde{\mathbf{A}}_{v_2} = 0$.

A similar argument constrains $\tilde{\mathbf{A}}_s$.  Again, because we expect that well established descriptions of 2D elasticity hold for developable bundles, we have that $\tilde{\mathbf{A}}_s$ should be independent of $s$ independent displacements $\mathbf{w}(v_1, v_2)$, as well as reptations, and that $\tilde{\mathbf{A}}_s \cdot D_{v_\alpha} \mathbf{r} = 0$.  This leaves $\tilde{\mathbf{A}}_s = c \mathbf{t}$, where $c$ is independent of the deformation.
Since, at the level of metric, $g^{\text{eff}}_{sI} = c^2 \delta_{sI}$ is independent of the deformation, no matter what $c$ is, it will not appear in the strain tensor, which has by definition of eq. (3) components only for $I \neq s$.  For simplicity, we take $\tilde{\mathbf{A}}_s = 0$.

%check this argument in the morning before sending along to g and c
%Now, while $\tilde{\mathbf{A}}_{I}$ can be almost any function of the deformation gradient $\nabla \mathbf{r}$, the argument preceding Eq.~(2) constrains it to transform as a scalar under reptations.  Taking $s = z$, $v_1 = x$, and $v_2 = y$ to be the usual Cartesian coordinates in $\mathbb{R}^3$, in which case $\tilde{\mathbf{A}}_I$ depends solely on the gradient of the 2D displacement field,
%\begin{equation}
%\mathbf{u} = u_1(s, v_1, v_2) \hat{x} + u_2(s, v_1, v_2) \hat{y}.
%\end{equation}
%However, the condition that a bundle is developable (again, that $\mathbf{t} = \hat{T}_0 (s)$) is impossible to determine without information from displacements along $\hat{z}$.  It is suffici, the deformation $\mathbf{r}(s,v_1,v_2) = [s + s g(v_1,v_2)] \hat{z} + [v_1 + s f(v_1,v_2)] \hat{x} + v_2 \hat{y} )$ is developable only when $f(v_1,v_2) = g(v_1,v_2) + 1 $).  So, $\tilde{\mathbf{A}}_{v_1} = \tilde{\mathbf{A}}_{v_2} = 0$ for any deformation.

\section{Small tilt limit}
We can recover the small-tilt approximation of the strain tensor from~\cite{selinger_hexagonal_1991,grason_chirality_2007,grason_defects_2012} starting from Eq.~(3), where the fully geometrically-nonlinear strain, $\epsilon_{\alpha \beta}$, is 
\begin{equation}
\epsilon_{\alpha \beta} = \tfrac{1}{2} \big [D_\alpha \mathbf{r} \cdot D_\beta \mathbf{r} - g_{\alpha \beta}^{\text{tar}} \big ].
\end{equation}
We break the deformation $\mathbf{r}$ up into the unstrained, cartesian coordinates $\mathbf{x}=x~\hat{x} +y~\hat{y} +z~\hat{z}  $ and a displacement field $\mathbf{u}$ in the $x-y$ plane, taking the arc-coordinate $s = z$.  Then,
\begin{equation}
D_\alpha \mathbf{r} \cdot D_\beta \mathbf{r} = \delta_{\alpha \beta} + \partial_\alpha u_\beta + \partial_\beta u_\alpha + \partial_\alpha \mathbf{u} \cdot \partial_\beta \mathbf{u} - (\mathbf{t} \cdot \partial_\alpha \mathbf{r}) (\mathbf{t} \cdot \partial_\beta \mathbf{r}),
\end{equation}
where $\mathbf{t} = (\hat{z} + \partial_s \mathbf{u})/\sqrt{1 + (\partial_s \mathbf{u})^2}$ and $u_\beta = \mathbf{u} \cdot (\partial_\beta \mathbf{x})$, and $\alpha, \beta = x, y$.  Subtracting off the Euclidean target metric, $g_{\alpha \beta}^{\text{tar}} = \delta_{\alpha \beta}$, we have
\begin{equation}
\epsilon_{\alpha \beta} = \tfrac{1}{2} \big [ \partial_\alpha u_\beta + \partial_\beta u_\alpha + \partial_\alpha \mathbf{u} \cdot \partial_\beta \mathbf{u} - (\mathbf{t} \cdot \partial_\alpha \mathbf{r}) (\mathbf{t} \cdot \partial_\beta \mathbf{r}) \big ].
\end{equation}
Now substituting for $\mathbf{t}$ and $\mathbf{r}$ in the last term, and grouping terms by power of the displacement field $\mathbf{u}$, we have:
\begin{align}
\epsilon_{\alpha \beta} &= \tfrac{1}{2} \big [ \partial_\alpha u_\beta + \partial_\beta u_\alpha + \partial_\alpha \mathbf{u} \cdot \partial_\beta \mathbf{u} - \frac{\partial_s u_\alpha \partial_s u_\beta}{1 + |\partial_s \mathbf{u}|^2} \nonumber \\
&- \frac{ \partial_s \mathbf{u} \cdot \partial_\alpha \mathbf{u} \partial_s u_\beta}{1 + |\partial_s \mathbf{u}|^2} - \frac{( \partial_s \mathbf{u} \cdot \partial_\beta \mathbf{u} ) \partial_s u_\alpha}{1 + |\partial_s \mathbf{u}|^2} - \frac{( \partial_s \mathbf{u} \cdot \partial_\alpha \mathbf{u} ) (\partial_s \mathbf{u} \cdot \partial_\beta \mathbf{u})}{1 + |\partial_s \mathbf{u}|^2}.
\end{align}
Expanding the denominator for small displacement fields, and keeping only terms which are quadratic in $\mathbf{u}$ recovers the rotationally invariant strain tensor,
\begin{equation}
\epsilon_{\alpha \beta}\simeq  \tfrac{1}{2} \big [ \partial_\alpha u_\beta + \partial_\beta u_\alpha + \partial_\alpha \mathbf{u} \cdot \partial_\beta \mathbf{u} - \partial_s u_\alpha \partial_s u_\beta \big].
\end{equation}

\section{Derivation of the force-balance equations}
We are looking for local extrema of the elastic energy
\begin{equation}
{\cal E}_s =\int dV ~ {\cal W} (\epsilon)   = \frac{1}{2} \int dV S^{\alpha \beta} \epsilon_{\alpha \beta},
\label{eqn:straineng}
\end{equation}
with respect to the deformation, $\mathbf{r}$, where $dV = ds dv^1 dv^2 \sqrt{\text{det}(g^{\text{tar}})}$, $g^{\text{tar}}$ is the target metric, as in Eq.~(3), $\epsilon_{\alpha \beta}$ is the strain tensor, as in Eq.~(3), and $S^{\alpha \beta}$ is the nominal stress tensor, as defined following Eq.~(5).  As such, we consider arbitrary variations $\delta \mathbf{r}$ of the energy around these local extrema, so that the restoring force on a small material volume is given by:
\begin{equation}
\frac{\delta {\cal E}_s}{\delta \mathbf{r}} = \int dV S^{\alpha \beta} \frac{\delta \epsilon_{\alpha \beta}}{\delta \mathbf{r}}.
\label{eqn:envar}
\end{equation}
What remains then is to work out $\frac{\delta \epsilon_{\alpha \beta}}{\delta \mathbf{r}}$, and apply the divergence theorem to derive the conditions of force balance. First, we note that $\epsilon_{\alpha \beta} = \tfrac{1}{2} \big [ D_\alpha \mathbf{r} \cdot D_\beta \mathbf{r} - g_{\alpha \beta}^{\text tar}],$ and that the covariant derivative $\nabla$ is just the usual partial derivative on scalars in the material space, so $\nabla_I \mathbf{r} = \partial_I \mathbf{r}$.
We then have
\begin{equation}
\delta \epsilon_{\alpha \beta}  = D_\beta \mathbf{r} \cdot \delta \big [ D_\alpha \mathbf{r} \big].
\end{equation}
Since $D_\beta \mathbf{r} \cdot \mathbf{t} = 0$, only two terms in $\delta \big [ D_\alpha \mathbf{r} \big ]$ contribute to $\delta \epsilon_{\alpha \beta}$:
\begin{equation}
\delta (\nabla_\alpha \mathbf{r}) = \delta (\partial_\alpha \mathbf{r}) =\partial_\alpha \delta \mathbf{r},
\end{equation}
and
\begin{equation}
(\mathbf{t} \cdot \nabla_\alpha \mathbf{r}) \delta \mathbf{t} =  (\mathbf{t} \cdot \nabla_\alpha \mathbf{r}) \frac{1}{|\partial_s \mathbf{r}|} \big [\partial_s \delta \mathbf{r} -   \mathbf{t} (\mathbf{t} \cdot \partial_s \delta \mathbf{r}) \big].
\end{equation}
All together, we have that
\begin{equation}
\delta \epsilon_{\alpha \beta}  = D_J \mathbf{r}\Big [\partial_\alpha \delta \mathbf{r} - \mathbf{t} \cdot \partial_\alpha \mathbf{r} \frac{\partial_{s} \delta \mathbf{r}}{|\partial_{s} \mathbf{r}|}\Big ],
\end{equation}
where again we use that $D_\beta \mathbf{r} \cdot \mathbf{t} = 0$.
Substituting this back into our integral, we find that
\begin{equation}
\delta {\cal E}_s = \int dV S^{\alpha \beta} D_\beta \mathbf{r} \cdot \Big [\partial_\alpha \delta \mathbf{r} - \mathbf{t} \cdot \nabla_\alpha \mathbf{r} \frac{\partial_{s} \delta \mathbf{r}}{|\partial_s \mathbf{r}|}\Big ].
\label{eqn:deltaE}
\end{equation}
Now we apply the divergence theorem, finding that
\begin{align}
0 &= - \int ds dv^1 dv^2 \partial_\alpha \Big [ \sqrt{\text{det}(g^{\text{tar}})} S^{\alpha \beta} D_\beta \mathbf{r} \Big ] \cdot \delta \mathbf{r} + \int ds dv^1 dv^2 \partial_\mathbf{s} \Big [\sqrt{\text{det}(g^{\text{tar}})} S^{\alpha \beta} D_\beta \mathbf{r} \frac{\mathbf{t} \cdot \partial_\alpha \mathbf{r}}{|\partial_s \mathbf{r}|} \Big ] \cdot \delta \mathbf{r} \label{eqn:elbulksi} \\
&+ \int dA~\hat{n}_\alpha \big [ \sqrt{\text{det}(g^{\text{tar}})} S^{\alpha \beta} D_\beta \mathbf{r} \big ] \cdot \delta \mathbf{r} - \int dA~\hat{n}_s \big [\sqrt{\text{det}(g^{\text{tar}})} S^{\alpha \beta} D_\beta \mathbf{r} \frac{\mathbf{t} \cdot \partial_\alpha \mathbf{r}}{|\partial_s \mathbf{r}|} \big ] \cdot \delta \mathbf{r}, \label{eqn:elboundarysi}
\end{align}
where $\hat{n}$ is the vector normal to the boundary of the material.  Eq.~\eqref{eqn:elboundarysi}, gives the boundary forces associated with residual stresses at the boundary, either directly from the stress tensor (first term), or from a non-zero flux of filament ends (second term).  Now noting that for a vector $f^I$ in the material space, $\partial_I \big ( \sqrt{\text{det}(g^{\text{tar}})} f^I ) = \sqrt{\text{det}(g^{\text{tar}})} \nabla_I f^I$, and that, since $\delta \mathbf{r}$ is an arbitrary variation, everything dotted into it must vanish, Eq.~\eqref{eqn:elbulksi} reduces to Eq~(5):
\begin{equation}
\label{eqn:el}
\frac{\delta \mathcal{E}_s}{\delta {\bf r} } =  -  \nabla_\alpha \big( S^{\alpha \beta} D_\beta \mathbf{r} \big )+  \nabla_s \Big( S^{\alpha \beta} D_\beta \mathbf{r} \frac{\mathbf{t} \cdot \nabla_\alpha \mathbf{r}}{|\nabla_s \mathbf{r}|} \Big ). 
\end{equation}

We obtain Eq~(6)--(8) by projecting Eq.~\eqref{eqn:el} along and perpendicular to $\mathbf{t}$.  The $\mathbf{t}$ component is
\begin{equation}
-    S^{\alpha \beta} \Big (\mathbf{t} \cdot \big [ \nabla_\alpha D_\beta \mathbf{r} \big ]\big )+    S^{\alpha \beta} \Big ( \mathbf{t} \cdot \big [ \nabla_s D_\beta \mathbf{r} \big ] \Big ) \frac{\mathbf{t} \cdot \nabla_\alpha \mathbf{r}}{|\nabla_s \mathbf{r}|}.
\end{equation}
Because $\mathbf{t} \cdot \nabla_\alpha D_\beta \mathbf{r} = - \partial_\alpha \mathbf{t} \cdot D_\beta \mathbf{r}$, $\mathbf{t} \cdot \nabla_s D_\beta \mathbf{r} = - \partial_s \mathbf{t} \cdot D_\beta \mathbf{r}$, and because the stress tensor is symmetric, we can rewrite this in terms of the {\it convective flow tensor} of the bundle,
\begin{equation}
h_{\alpha \beta} = \tfrac{1}{2} \Big [ \partial_\alpha \mathbf{t} \cdot \partial_\beta \mathbf{r} + \partial_\beta \mathbf{t} \cdot \partial_\alpha \mathbf{r} -  \frac{\mathbf{t} \cdot \partial_\alpha \mathbf{r}}{|\partial_s \mathbf{r}|}\partial_s \mathbf{t} \cdot \partial_\beta \mathbf{r} - \frac{\mathbf{t} \cdot \partial_\beta \mathbf{r}}{|\partial_s \mathbf{r}|}\partial_s \mathbf{t} \cdot \partial_\alpha \mathbf{r} \Big ],
\end{equation}
as
\begin{equation}
F_\parallel = S^{\alpha \beta} h_{\alpha \beta},
\label{eqn:betteryetelbulkpar}
\end{equation}
recovering Eq.~(6).

The orthogonal forces follow straightforwardly by projecting out the tangent component and the definition of the gauge covariant derivative, as  $D_\alpha \mathbf{F}^{\alpha} = \nabla_\alpha \mathbf{F}^{\alpha} - (\mathbf{t} \cdot \nabla_\alpha \mathbf{F}^{\alpha}) \mathbf{t}$, from which we recover Eq.~(8).

\section{The energetics of twisted-toroidal bundles with small curvatures}
From Eqs.~(6) and~(8), we can solve for the stable configurations of bundles of helices with constant pitch $2 \pi / \Omega$ (i.e., the twist axis is unbent).  For a hexagonal columnar phase, the tensor of elastic constants in the material frame is
\begin{equation}
C^{\alpha \beta \gamma \delta} = \frac{ Y}{1+\nu} \Big [ \frac{\nu}{(1-\nu)} g_{\text{tar}}^{\alpha \beta}g_{\text{tar}}^{\gamma \delta}  + \tfrac{1}{2}\big ( g_{\text{tar}}^{\alpha \gamma} g_{\text{tar}}^{\beta \delta} + g_{\text{tar}}^{\alpha \delta} g_{\text{tar}}^{\beta \gamma} \big ) \Big ],
\end{equation}
where $Y$ and $\nu$ are the 2D Young's modulus and Poisson ratio, respectively.
%Need to rework this a little to smooth over coordinate confusion!
The deformation field for a bundle of helices with uniform pitch is
\begin{equation}
\mathbf{r}^{(0)} = \mathbf{r}_0(s) + \rho(r) \hat{\rho}(\phi),
\label{eqn:helicaldef}
\end{equation}
where here $\mathbf{r}_0(s)$ is a central reference curve which we take to be a straight line with $\partial_s^2 \mathbf{r}_0 =0$, $\hat{\rho}$ is the usual radial unit vector in cylindrical coordinates in the target space, $\phi = \varphi + \Omega s$, and $r$, $\varphi$, and $s$ are the radial, polar, and cylyndrical coordinates in the material space.  The tangent field then lies along $\partial_s\mathbf{r}^{(0)} = {\bf t}_0 + \Omega \rho \hat{\phi}$, so $\mathbf{t}^{(0)} =  \big ( {\bf t}_0 + \Omega \rho \hat{\phi}\big )/\sqrt{1+\Omega^2\rho^2}$.
In the absence of external forces, Eqs.~(6) and~(8) now reduce to a nonlinear boundary value problem (BVP) for $\rho(r)$:
\begin{align}
0 &= - \partial_r \Big \{ r \rho'(r) \big [ \frac{Y}{2-2\nu^2} (\rho'(r)^2 - 1) + \frac{\nu Y}{2-2\nu^2} (\frac{\rho(r)^2}{r^2 (1+\Omega^2 \rho(r)^2)} - 1 ) \big ] \Big \} \nonumber \\
&- \frac{\rho(r)}{r (1+ \Omega^2 \rho(r)^2)^2} \big [\frac{\nu Y}{2-2\nu^2} (\rho'(r)^2 - 1) + \frac{Y}{2-2\nu^2} (\frac{\rho(r)^2}{r^2 (1+\Omega^2 \rho(r)^2)} - 1 ) \big ] \label{eqn:helicalbvp}\\
0 &= \rho(0) \\
0 &= \frac{Y}{2 - 2 \nu^2} (\rho'(R)^2 - 1) + \frac{\nu Y}{2 - 2 \nu^2} \Big( \frac{\rho(R)^2}{R^2 (1+ \Omega^2 \rho(r)^2)} - 1 \Big ).
\end{align}
Solving this BVP numerically gives the equilibrium deformation field for the straight helical bundle, as shown in Fig.~1.

%Gotta find a way to describe this here deformation! We're doing something a little weird in the perturbation (\partial_s \mathbf{t}_0 \mathcal{O}(\epsilon)), but uhh that shouldn't stop us from trying to make it clear!
To find the low energy configurations of weakly curved twisted-toroidal bundles, we now introduce a perturbation at $\mathcal{O}(\kappa_0)$ to Eq.~\eqref{eqn:helicaldef} by taking $\partial_s^2 \mathbf{r}_0 = \kappa_0 \hat{N}$ and $\mathbf{r} = \mathbf{r}^{(0)} + \mathbf{r}^{(1)} + \mathcal{O}(\kappa_0^2)$, with
\begin{equation}
\mathbf{r}^{(1)} = \delta s \partial_s \mathbf{r}^{(0)} + \delta \rho \hat{\rho} + \delta \phi \hat{\phi},
\end{equation}
and expand the elastic energy in Eq.~(4) to quadratic order in $\kappa_0$.  The linear correction to the elastic energy vanishes because the constant pitch helical bundles are in force balance, and so the resulting elastic energy takes the form
\begin{equation}
\mathcal{E}_s = \mathcal{E}_{\text{helical}} + \tfrac{1}{2} \kappa_0^2 \mathcal{E}_{\text{correction}}[\delta s, \delta \rho, \delta \phi].
\end{equation}
Scaling analysis of the force balance equations shows that, at small $\rho$, the components of $\mathbf{r}^{(1)}$ are
\begin{align}
\delta s &= a_s \Omega \kappa_0 \rho^3 \sin{\phi} \nonumber \\
\delta \rho &= a_\rho \Omega^2 \kappa_0 \rho^4 \cos{\phi} \label{eqn:ansatzdefsi} \\
\delta \phi &=a_\phi \Omega^2 \kappa_0 \rho^3 \sin{\phi}, \nonumber
\end{align}
as in Eq.~(9).  To stop the bundle from unbending and effectively decreasing its curvature, we fix the average position along the normal vector, $\hat{N}$, at $\mathbf{r}_0$, so that
\begin{equation}
\frac{1}{A} \int dA \hat{N} \cdot \mathbf{r}^{(1)} = 0.
\end{equation}
%Similar coordinate confusion problem here.
This constrains $a_s = a_\rho - a_\phi$, since
\begin{equation}
\frac{1}{A} \int dA \hat{N} \cdot \mathbf{r}^{(1)} = - (a_s - a_\rho + a_\phi) \frac{\Omega^2 \kappa_0}{R^2} \int_0^R dr r \rho(r)^4.
\end{equation}
Having found $\rho(r)$ from Eq.~\eqref{eqn:helicalbvp}, we can subtitute the ansatz from Eq.~(9) into the elastic energy in Eq.~(4) and integrate over the volume for a given Young's modulus, $Y$, 2D Poisson's ratio, $\nu$, and reciprocal pitch, $\Omega R$, to obtain an elastic energy
\begin{equation}
\mathcal{E}_s = \mathcal{E}_{\text{helical}} + \tfrac{1}{2} \kappa_0^2 \mathcal{E}_{\text{correction}}(a_\rho, a_\phi).
\label{eqn:quadraticcorrection}
\end{equation}
Eq~\eqref{eqn:quadraticcorrection} is quadratic in $a_\rho$ and $a_\phi$ and has a minima at $a_\rho$, $a_\phi \neq 0$ whenever $\Omega R$ and $\kappa R$ are nonzero.  These energy minimizing displacements, and the resultant pressure in the cross-section, are shown in Fig.~1 for a given $\kappa_0 R$.  The resultant elastic energy per unit length takes the form of an effective bending modulus, as shown Fig.~3.

Twisted toroidal bundles are non-equidistant, and we can calculate the components of the convective flow tensor from the perturbative displacements in Eq.~\eqref{eqn:ansatzdefsi}.  Since the uniform pitch helices are equidistant, the leading contribution to the convective flow tensor is linear in $\kappa_0$:
\begin{align}
h^{(1)}_{\rho \rho} &= 
\frac{ \rho'(r) \partial_s \partial_r \delta \rho}{\sqrt{1+\Omega^2 \rho^2}} \nonumber \\
h^{(1)}_{\rho \phi} &=  \frac{\rho^2 \partial_s \partial_r \delta \phi + \rho'(r) (1+\Omega^2 \rho^2)\partial_s \partial_\varphi \delta \rho - \Omega \rho^2 \rho'(r) \partial_s^2 \delta \rho}{2 (1+\Omega^2\rho^2)^{3/2}}\\
h^{(1)}_{\phi \phi} &=  \frac{\Omega^3 \kappa_0 \rho^5 + \rho \partial_s \delta \rho + \rho^2(1+\Omega^2\rho^2) \partial_s \partial_\varphi \delta \phi - \Omega \rho^3 \partial_s^2 \delta \phi}{(1+\Omega^2 \rho^2)^{5/2}}. \nonumber
\end{align}
By substituting Eq~\eqref{eqn:ansatzdefsi} into the above expression for $h_{\alpha \beta}$, and using the lowest energy values of $a_\rho$ and $a_\phi$, we can calculate the elasticity-mediated response of twisted-toroidal bundles to the geometric constraints on constant spacing.
The contributions to non-equidistance can be broken up into the splay, $ \text{tr}(h)$, and biaxial splay, $ \big (h_{\alpha \beta} - \tfrac{1}{2} \text{tr}(h) \delta_{\alpha \beta} \big ) = h_{\text{dev}}$, of the tangent field, $\mathbf{t}$.  To measure the contributions to these two modes from the linear displacements driven by elastic interactions of twisted-toroidal bundles, we compare their integrated, dedimensionalized contributions to the Frank free energy,
\begin{align}
\Gamma &= \frac{1}{\kappa_0^2 V} \int dV \text{tr}(h_{\text{dev}}^2) \\
\Psi &= \frac{1}{\kappa_0^2 V} \int dV \text{tr}(h)^2,
\end{align}
where $dV = ds dv^1 dv^2 \sqrt{\text{det}(g^{\text{tar}})}$, as shown in the inset of Fig.~3.

\end{document}